# A Blockchain based Fund Management System for Construction Projects
## A Comprehensive Case Study in Xiong'an New Area China

Wenlue Song[1], Hanyuan Wu[2], Hongwei Meng[2], Evan Bian[2], Cong Tang[3], Jiaqi Xi[4], Haogang Zhu[2]
[1.] Beihang University  [2.] Xiong'an Blockchain Lab, China  [3.] Peking University  [4.] Beijing University of Technology
songwenlue@buaa.edu.cn, {menghongwei, bianyifu, wuhanyuan}@xaicif.org.cn,
tangcong@infosec.pku.edu, jiaqi.xi@emails.bjut.edu.cn

*Abstract* - **As large-scale construction projects become increasingly complex, the use and integration of advanced technologies are being emphasized more and more. However, the construction industry often lags behind most industries in the application of digital technologies. In recent years, a decentralized, peer-to-peer blockchain technology has attracted widespread attention from academia and industry. This paper provides a solution that combines blockchain technology with construction project fund management. The system involves participants such as the owner's unit, construction companies, government departments, banks, etc., adopting the technical architecture of the "Xiong'an Blockchain Underlying System". The core business and key logic processing are all implemented through smart contracts, ensuring the transparency and traceability of the fund payment process. The goal of ensuring investment quality, standardizing investment behavior, and strengthening cost control is achieved through blockchain technology. The application of this system in the management of Xiong'an construction projects has verified that blockchain technology plays a significant positive role in strengthening fund management, enhancing fund supervision, and ensuring fund safety in the construction process of engineering projects. It helps to eliminate the common problems of multi-party trust and transparent supervision in the industry and can further improve the investment benefits of government investment projects and improve the management system and operation mechanism of investment projects.**

## 1. INTRODUCTION

The significant characteristics of construction projects are large funds, high risk, and long construction periods, and the complexity of their construction process is unprecedented. At present, some problems that cannot be ignored are exposed in the financial and fund management of engineering projects, especially in the aspects of multi-party cooperation, trust, information sharing, process automation, transparency, and supervision, facing many challenges [1]. Therefore, it is urgent to improve the management capabilities of construction projects by adopting new technologies and mechanisms.

In recent years, digitalization is crucial in improving industry efficiency, and the application of blockchain technology in various fields of society has once again attracted people's attention [2]. Some researchers believe that blockchain technology may completely change the engineering industry by changing its production and procurement methods [3], and it has the potential to change many industries around the world, including the construction industry [4].

However, at present, the construction industry is slow in the application of digital technologies. In the past 50 years, the efficiency improvement of the construction industry has only been half of other industries [5]. According to a global report released by McKinsey, during the "Industry 3.0" period, the use of information technology in the construction industry ranked second to last [6]. With the gradual maturity of blockchain technology, blockchain has brought new opportunities for construction management [7]. Whether blockchain can lead a digital revolution as it does in other industries remains to be actively explored and applied [8][9].

Therefore, how to strengthen fund management, strengthen fund supervision, and ensure the safety and integrity of funds is one of the cores of project management. To solve the above problems, by establishing a blockchain-based construction fund management information system, breaking the information island, realizing data sharing, improving approval efficiency, and improving investment benefits. The system has the following characteristics:

1) The blockchain architecture of this project is based on the Xiong'an's independent and controllable blockchain underlying system, adopts the architecture of urban blockchain layered multi-chain, has the technical characteristics of security and control, and openness and compatibility, and lays a solid foundation for subsequent cross-chain interaction and multi-scenario applications.

2) All data fields strictly follow the Xiong'an New Area's engineering big data governance standards. It provides basic conditions for subsequent data integration and data open sharing.

3) The system design adopts the technical idea of DApp (decentralized application), and the business process and key logic processing are all realized through smart contracts. The key data is directly obtained from the chain, realizing the approval flow driving the capital flow, ensuring the transparency and traceability of the payment process. At the same time, in the process of interacting with the blockchain, it follows the principle of who initiates and who signs.

## 2. BACKGROUND AND RELATED WORK

Blockchain is a concept derived from Bitcoin. It is a chain structure of data blocks arranged in chronological order. In essence, it is a distributed database that realizes the security and tamper-proof of each link in a decentralized manner and combined with cryptography [10]. The white paper [11] proposes the steps to run a blockchain network. In order to promote the blockchain industry to accelerate technological innovation and application landing, the Information Center of the Ministry of Industry and Information Technology released the "2018 China Blockchain Industry White Paper"[12], which deeply analyzes the application scenarios of blockchain technology in the financial field and the real economy in China, and systematically expounds the six characteristics and ten trends of the development of China's blockchain industry. [13] implemented a road-side parking management system based on the Hyperledger Fabric permissioned chain, collecting parking information through edge devices and storing key data on the blockchain platform. [14] used the VNT Chain blockchain platform to design and build a digital music copyright management system, using blockchain technology to provide evidentiary proof for music copyrights and to realize evidence solidification to provide originality proof. [15] studied a precision poverty alleviation data protection scheme based on blockchain. Relying on smart contracts and interstellar file systems and other technologies, data is added, updated, verified, and shared in the form of digital files. [16] proposed a time-dimension medical data security sharing scheme based on the consortium chain. The scheme can securely store medical data while realizing fine-grained access control with a time dimension. [17] proposed a lightweight blockchain architecture suitable for task scheduling in the vehicle-connected cloud, using roadside units to build a blockchain network, and adopting an improved practical Byzantine fault tolerance algorithm to complete consensus. In the field of construction, [18] proposed three applications of blockchain in construction management, suggesting that blockchain can solve the problem of insufficient trustworthy information resources in prefabricated supply chain management, but there is little research on the application of blockchain in the construction industry. In response to the challenges often faced in supply chain management, such as fragmentation, poor traceability, and lack of real-time information, [19] researched and established a blockchain-based supply chain information management framework, extending the application of blockchain in the field of construction supply chain. [20], based on the analysis of the basic model of blockchain technology, adopted the "consortium chain + private chain" dual-chain smart contract, and established an intelligent water conservancy information sharing platform including government departments, water conservancy enterprises, the public, and third-party maintenance. [21] discussed the risks and applications of blockchain technology in improving end-to-end design and construction processes. The results of case studies show that the deployment of blockchain can help companies save 8.3% of the total construction cost and pointed out the future research direction of blockchain technology in construction engineering. In [22], in order to support engineering quality management, explored a construction quality information management framework based on blockchain, detailed the consensus process to solve the problem of information fraud, extended the application of blockchain in the field of construction management, and verified the feasibility of the system.

## 3. SYSTEM REQUIREMENTS AND GOALS

Under the traditional management mode, the project progress information of the construction and management system has not been connected with the fund management action, and the project progress is out of line with the fund payment, and the fund allocation needs to be paid separately. The data of the construction and management system is not perfect enough, the progress information is not detailed enough, and the contract information is not standardized. The use of construction funds layer upon layer of approval, manual payment consumes energy, there is no intelligent allocation. Project funds are allocated layer by layer, there may be interception, misappropriation and arrears, and there is no full link management of funds.

By building a construction fund management information system with rigorous audit and transparent flow direction, Solution designed to provide a true blockchain fund management system, With the blockchain as the underlying logic, Smart contracts trigger all kinds of transactions, The payment process requires no human intervention, Build a decentralized application thoroughly, To truly realize the automatic payment of funds through the blockchain, Improve the timeliness and accuracy of the distribution of construction project funds to suppliers and labor service personnel, Effectively promote the sound development of the supply chain, Help the government to solve various project management and fund management problems, Effectively guarantee the stable operation of urban construction projects, Further enhance the intelligence and modernization of urban governance and management.

## 4. SYSTEM DESIGN

*4.1 System business design*

The applicable user groups of this project are government departments, owners (project construction units), general contractors, subcontractors, raw material suppliers, labor service companies, etc. The basic business process of the system design is divided into two parts: project management and payment. The details are shown in Figure 1. It mainly involves the following parts of the business development:

1) According to the requirements, on the developed blockchain platform of multi-chain coexistence, cross-chain interconnection and double-layer operation, the core chain is used to connect cross-chain multiple payment chains, realizing the data convergence of multiple chains and realizing the major innovation of urban blockchain.

2) Project management, to realize the whole process of government supervision. By pooling the information of all parties and changing the payment to blockchain based, the whole process of the project can be supervised.

3) Through the payment, to achieve the transparent and timely distribution of construction funds. Through the platform, the construction funds can be allocated through, and the funds will be directly allocated from the project account to the subcontractors / suppliers, so as to improve the efficiency of fund transfer and ensure the accurate and timely fund allocation. Through the platform, the wages are directly allocated to the personal bank account of the labor personnel to realize the timely and accurate payment of the labor personnel.

4) Create a dashboard to help the government coordinate the management. Through the platform timely and accurate access to project budget, construction schedule, funds allocated, etc., and through the system to form all kinds of statistics, finally achieve the effect of the project life cycle management, so that the government can timely grasp the engineering funds allocated information, realize the government of all kinds of project progress control and use of funds.

5) Key management to ensure the security of the platform fund allocation. Through the platform, during the submission of payment application and payment review, redefine the identity of the operator, reduce the risk of platform fund allocation, realize the verification of the key management plug-in under relevant high-risk operation scenarios, and reduce the risk of fund allocation during the payment operation.

*4.2 System architecture design*
*4.2.1 Logical architecture*

Among the users of the platform, government, enterprise and bank personnel can access WEB applications through the Internet for data processing and information interaction. The customized payment adaptation layer is connected with several banking systems, to solve the inconsistent interfaces of each banking system due to personalized reasons, and to obtain the batch distribution service of funds. Project management module, organization management module and fund management module belong to the core business, which is used to manage the project situation and project fund distribution, guarantee the project progress and supervise the fund dynamics. There are some auxiliary modules: (1) provide system management functions to manage users, roles and rights; (2) provide workbench to manage workflow; (3) provide cockpit and report statistics for the decision-making level to view the overall progress of the project and fund distribution, and control the overall situation. The blockchain platform provides smart contract services for business process processing, as well as storage fragmentation and data expansion services, for access to the third-party blockchain platform. The data center obtains the information on the chain through the SDK, transforms it into a relational data structure, and provides the query service for the WEB applications. The detailed logical architecture of the system is shown in Figure 1.

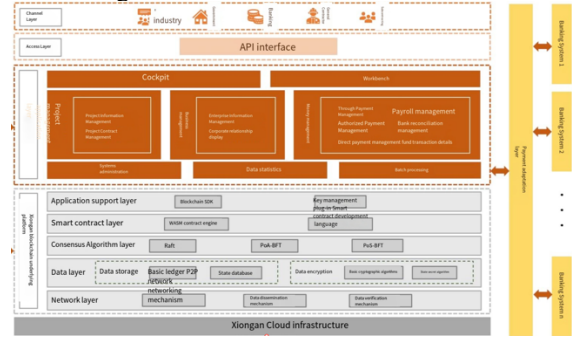

Figure 1. System logic architecture

*4.2.2 Physical architecture*

The physical architecture of the system is shown in Figure 2. All the servers are deployed in clusters to meet the high availability requirements. Shared storage servers are used to store data such as contracts. Blockchain will have more than 4 parties participating, with each party having 2 PEER nodes. Personal key management tool directly interacts with the blockchain platform through the SDK. All nodes are deployed in Xiongan Cloud, and restricted external ports are used to reduce the exposure risk. Use various security services provided by Xiongan Cloud to reduce risks, such as: WEB application firewall, DDOS traffic cleaning and other services.

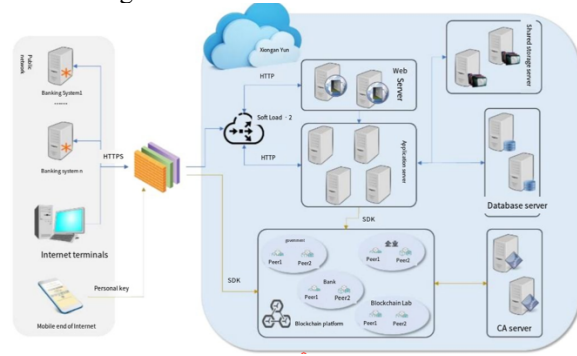

Figure 2. Physical architecture of the system

*4.3 System module design*

The system consists of cockpit, workbench, project management, contract management, supply chain management, fund management, system management, data report, and blockchain service module, as shown in Figure 3.

1) The Dashboard is visually displayed to view the key data and statistics.

2) The workbench is responsible for distributing, managing and tracking the tasks in the contract and payment approval process.

3) The project management module is responsible for the unified input and management of all levels of projects in the new district by stages. The write permission of this module is owned by system-level users, and the read permission is owned by system-level users and associated institutional users. The whole project management process is mainly divided into five stages: feasibility study, initial design, budget, construction and completion.

4) Contract management is responsible for maintaining the project contract list, and is also responsible for the contract creation, modification, delete, search, view, review and other management functions.

5) Supply chain management is based on the addition, deletion, modification and investigation of the owner, general contractor and subcontracting organizations involved in the project. The system administrator is responsible for maintaining the basic information of the organization, and the organization administrator is responsible for maintaining the organization's bank account information.

6) Fund management is responsible for the management and maintenance of various payment scenario processes and data covered by application payment, authorized payment and penetrating payment.

7) System management mainly includes the relevant users, roles and authority management in the system.

8) Data report is responsible for the integration, analysis and statistical report display and export of all kinds of data in the system.

9) Blockchain services include blockchain payment services, user key generation and management, and blockchain data middle platform services.

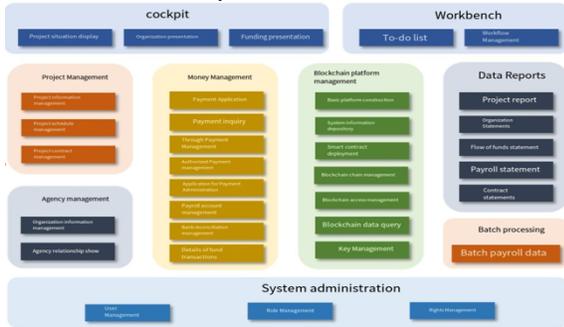

Figure 3. System function module

## 5. SYSTEM TEST AND APPLICATION

*5.1 Test environment*

The operating system is Ubuntu 16.04, the processor is 8C, the memory is 16G, and the hard disk storage is greater than 500G. The KeepAlived 1.2.19 is used to build the load balancing system with IPVS, the Web server uses Nginx 1.12.1, the database is Mysql Innodb5.7.29, and the shared storage service uses Glusterfs 7.5.1 distributed file system. Response time performance test, interface test use YAPI, and performance test use Loadrunner, AppScan and Network Mapper for safety test. The test data used are the simulated data constructed according to the real data structure of the system during its actual use.

*5.2 Functional test*

In the Ubuntu 16.04.6 LTS system test environment, all the functional test points of Xiongan Construction fund blockchain platform project, the payment process test fully covers 3 modules and 8 scenarios in the fund management, a total of 68 execution cases, and the execution of each function case is normal. Each business side process can meet the needs of users, accurate data, friendly user interface, easy to operate, stable interface and correct database storage, in line with user operation habits, can be compatible with common web browsers and different resolutions.

*5.3 Integration testing*

Ensure the correctness of the interface between the fund management platform and the blockchain platform, the correctness of the interface between the blockchain platform and the bank adaptation service, the correctness of the bank adaptation service and the bank-enterprise interconnection interface, and the connectivity test of the fund management platform, the blockchain, the bank adaptation service and the bank-enterprise side. The test scope covers the test fund management platform, blockchain platform, bank adaptation services and bank-enterprise interconnection interface. The test process is shown in Figure 4.

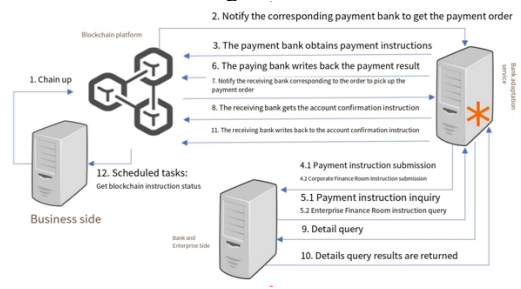

Figure 4. Integration test

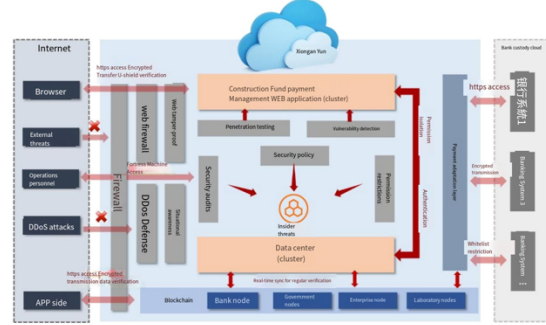

Figure 5. System safety test

*5.4 Safety test*

The overall structure and security scheme of the system boundary access platform are shown in Figure 5. The goal of safety test is to find the possible safety risks in the current system and avoid hazardous safety events. For the security of the website and web system, the following aspects are tested: Web server security vulnerability, Web server error configuration, SQL injection, XSS (cross-site script), directory traversal, input authentication, password protection vulnerable permission directory, dangerous HTTP method, security request path check.

The security test tool AppScan is used for security scanning test. Security audit test was performed using Network Mapper. Identify all the major weaknesses that cyber attackers can exploit.

## 5.5 Performance test

Through the performance test of the system, we can understand the maximum concurrent capability supported by the construction fund management platform and estimate the business capacity of the system.

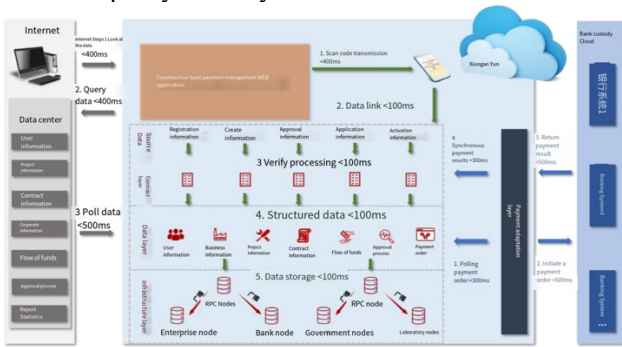

Figure 6. Performance test

Through typical transaction selection, the test transaction functions include: login, project management-> project management list, contract management-> contract list, contract management-> create contract, fund management-> payment application (8 scenarios: supplier materials please, please pay, builders, project advance payment, project progress payment, project payment, builder wages, daily reimbursement, staff loan issue), fund management-> application list, fund management-> pay to state query, view a total of 13 functions.

The test and implementation of various business scenarios provide data reference for system tuning. Understand the stability of the business system, focus on memory leakage, data increase response time, downtime and other problems. Through the performance test, the system bottleneck is found in time, and the development and debugging are coordinated to achieve the optimal purpose of the system to test whether the operating status of the system in various scenarios, the maximum throughput of the system and the processing capacity of the system reach the predetermined indexes. The system call and response process are shown in Figure 6.

## 5.6 System Application

As of May 2023, 426 government investment bidding projects in Xiongan New Area are running on the chain, involving more than 4,600 enterprises such as owners, general contractors and subcontractors, with a total of more than 9,500 contracts on the chain, with a cumulative payment amount of 57.1 billion yuan, of which 43.25 million Xiongan construction workers' wages are issued, and the model has gained double recognition from the government + market, using the traceability characteristics of blockchain to establish a target tracking mechanism for financial funds and state-owned assets The model has achieved dual government + market recognition, using blockchain traceability to establish a targeted tracking mechanism for financial funds and state-owned assets, and implementing the social benefits of financial funds.